\documentclass[conference]{IEEEtran}
\IEEEoverridecommandlockouts
\usepackage{amsmath,amsfonts,amssymb}
\usepackage{stmaryrd}
\usepackage{bbm,bm}
\usepackage{textcomp}
\usepackage{xcolor}
\usepackage{soul}
\usepackage[nameinlink]{cleveref} 
\usepackage{graphicx}

\usepackage[letterpaper]{geometry}
\usepackage{todonotes}
\usepackage{url}

\usepackage[export]{adjustbox}  

\usepackage{algorithm2e}  
\usepackage{amsthm}
\theoremstyle{plain}  

\newcommand{\ie}{\emph{i.e.}}

\newcommand{\eg}{\emph{e.g.}}
\newcommand{\UCB}[0]{$\mathrm{UCB}$}

\makeatletter
\newcommand{\removelatexerror}{\let\@latex@error\@gobble}  
\makeatother

\title{
	Upper-Confidence Bound for Channel Selection in LPWA Networks with Retransmissions
	\thanks{
		This publication is supported by
		the French National Research Agency (ANR), under the projects SOGREEN and EPHYL (grants \emph{N ANR-14-CE28-0025-02} and \emph{N ANR-16-CE25-0002-03}), by R\'egion Bretagne, France,
		by \'Ecole Normale Sup\'erieure de Paris-Saclay.
		by European Union, through the European Regional Development Fund (ERDF),
		and by Ministry of Higher Education and Research, Brittany and Rennes Métropole, through the CPER Project \emph{SOPHIE / STIC \& Ondes}.
	}
}

\author{
	R{\'e}mi Bonnefoi$^{1}$, Lilian Besson$^{1}$, Julio Manco-Vasquez$^{1}$, and Christophe Moy$^{2}$ \vspace{0.3cm} \\
	$^1$ IETR / CentraleSup{\'e}lec Campus de Rennes, F-$35510$ Cesson-S{\'e}vign{\'e}, France, \\
	\texttt{$\{$Remi.Bonnefoi,Lilian.Besson,JulioCesar.MancoVasquez$\}$}\texttt{{@}CentraleSupelec.fr} \\
	$^2$ Univ Rennes, CNRS, IETR - UMR $6164$, F-$35000$, Rennes, France \\
	\texttt{Christophe.Moy{@}Univ-Rennes1.fr}
}

\begin{document}

\maketitle

\begin{abstract}
	In this paper, we propose and evaluate different learning strategies based on Multi-Arm Bandit (MAB) algorithms. They allow Internet of Things (IoT) devices to improve their access to the network and their autonomy, while taking into account the impact of encountered radio collisions.
	For that end, several heuristics employing Upper-Confident Bound (\UCB{}) algorithms are examined, to explore the contextual information provided by the number of retransmissions.
	Our results show that approaches based on \UCB{} obtain a significant improvement in terms of successful transmission probabilities.
	Furthermore, it also reveals that a pure \UCB{} channel access is as efficient as more sophisticated learning strategies.
\end{abstract}

\begin{IEEEkeywords}
	Low Power Wide Area, Multi-Armed Bandits, Upper-Confident Bound, retransmissions, Internet of Things.
\end{IEEEkeywords}

\section{Introduction}\label{seq:introduction}

Nowadays, the Internet of Things (IoT) and in particular the Low Power Wide Area (LPWA) technology is considered a main driver for a vast variety of application that will support the communications among a large number of devices.
In fact, network operators are starting to deploy Machine to Machine (M2M) solutions using LPWA networking technologies \cite{Raza17}.
For instance, LoRaWAN and SigFox technologies have been most adopted in the monitoring of large scale systems (\eg, smart cities, metering), where a large number of devices compete for the transmission of their packets in the unlicensed Industrial, Scientific and Medical (ISM) bands.

Nevertheless, this demand to fit a growing number of energy-limited end-devices requires the development of contention-based protocol more tailored for LPWAN technologies.
Thus, novel access mechanisms considering collision-avoidance methods need to be addressed to avoid degrading the network performance in these unlicensed bands.
In fact, the number of packet collisions increases as more devices without coordination share the same band.
Hence, an important concern in the  Medium Access (MAC) design is to reduce the Packet Loss Ratio (PLR) due to the interference caused by the collisions among the devices.

In this regard, in the context of Cognitive Radio \cite{Mitola99,Haykin05},
Multi-Arm Bandit (MAB) algorithms \cite{Auer,Auer02,Bubeck12} have been recently proposed as a potential solution for channel access in LPWA networks \cite{Bonnefoi18,Azari18,Bonnefoi17}.
For instance in \cite{Bonnefoi17}, the impact of non-stationarity on the network performance using MAB algorithms is studied.
In this work, low-cost algorithms following two well-known approaches, such as the Upper-Confidence Bound (\UCB{}) \cite{Auer,Auer02}, and the Thompson Sampling (TS) algorithms \cite{Thompson33} have reported encouraging results.
Other recent directions include theoretical analysis \cite{BessonALT18,BoursierPerchet18},
and realistic empirical simulations \cite{kumar2016two,kumar2017channel},
of the application of MAB algorithms for slotted wireless protocols in a decentralized manner,
or applications to multi-hoping networks \cite{Mitton,Toldov}. None of the above mentioned articles discusses in detail the impact of retransmissions on the performance of MAB learning algorithms as we do in this paper.

The aim of this paper is to assess the performance of MAB algorithms \cite{Bubeck12} for channel selection in LPWA networks, while taking into account the impact of retransmissions on the network performance.
For this reason, several decision making strategies are applied after a first retransmission (\ie, when a collision occurs).
Proposed approach employs contextual information provided by the number of retransmissions, and implemented at each device, so that no coordination among them is needed.
Moreover, our \UCB{}-based heuristics show low complexity making them suitable for being embedded in LPWA devices.

The contributions of this paper are summarized as follows:
\begin{itemize}
	\item
	Firstly, we provide a close form approximation of the radio collision probability after a first retransmission.
	By doing this, we highlight the need to develop a learning approach for channel selection upon collision.

	\item
	Secondly, different heuristics are proposed to cope with retransmissions.

	\item Lastly, we conduct simulations in order to compare the performance of the proposed heuristics with a naive uniform random approach, and a \UCB{} strategy (\ie, without any learning for the retransmissions).
\end{itemize}

The rest of the paper is organized as follows.
First the system model is introduced in Section~\ref{sec:model}.
Our motivations are exposed in Section~\ref{sec:motivations}, and a formal description of the MAB learning algorithms is given in Section~\ref{sec:MABalgo}. The proposed \UCB-based heuristics are presented in Section~\ref{sec:heuristics}, while the corresponding numerical results are shown in Section~\ref{sec:numExp}. Finally, some conclusions are drawn in Section~\ref{sec:conclusion}.

\section{System model}\label{sec:model}

\subsection{LPWA Network}

We consider in this paper an LPWA network composed of a gateway and a large number of end-devices that regularly send short data packets, where $K$ channels ($K>1$) are available for the transmission of their packets.

We assume that this network is constituted by two types of devices:
on one hand, we have \emph{static} devices that operate in one channel\footnote{~Note that, for unlicensed bands, this definition also encompasses any device following a different standard or trying to establish communication with gateways of other networks.} in order to communicate with the gateway.
On the other hand, there are  IoT devices, that possess the additional advantage of being able to select any of the $K$ available channels to perform their transmissions.

Regardless the type of devices, each of them follows a slotted ALOHA protocol \cite{Roberts75}, and has a probability $p>0$ to transmit a packet in a time slot.
We make the hypothesis that the transmission is successful if the channel is available, otherwise upon radio collision, these devices will attempt to transmit their packet up-to $M$ times, with $M \in\mathbb{N}$.
Note that, every retransmission is carried out after a random back-off time, uniformly distributed in $\llbracket 0, m-1 \rrbracket$, where $m \in\mathbb{N}, m>0$ is the length of the back-off interval.

\subsection{Model of our IoT devices}

\begin{figure}[htp!]  
	\centering
	\includegraphics[width=1.00\linewidth]{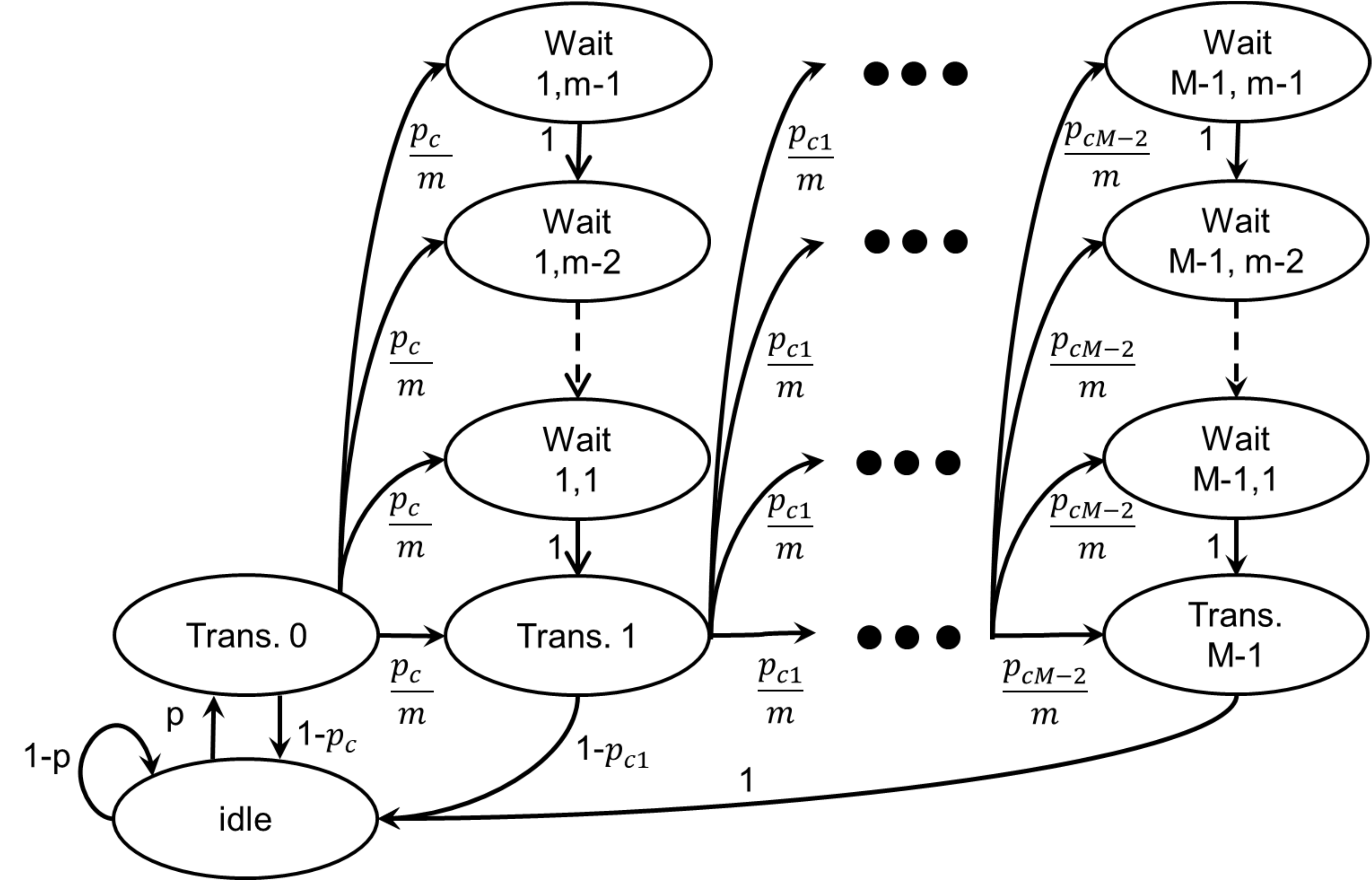}
    \caption{All devices in the network follow the same Markov behavior.}
	\label{fig:Markov_model}
\end{figure}

The aforementioned contention process can be described by a Markov chain model \cite{Norris98} similar to the one presented in \cite{Yang12}, as it is depicted in Fig.~\ref{fig:Markov_model}.
A device containing a packet for transmission goes from an idle state to a transmission state, while considering retransmissions due to different collision probabilities, \ie, $\{p_{c}, p_{c1}, \dots, p_{cM-2} \}$, at each $M-1$ back-off stage.
At each time slot, a transition from an idle state to a transmission state (denoted as \texttt{Trans.}) occurs if a packet transmission is required, while waiting states (denoted as \texttt{Wait}), correspond to a $m$ back-off interval.

A device aims to select a channel with the highest probability of successful transmission, for which it resorts to a reinforcement learning approach. It is formulated as a MAB problem, where each channel (also called arms) is viewed as a gambling machine (bandit), and each bandit has a \emph{reward}. Then, at every trial, a device chooses a channel that maximizes the sum of the collected rewards. These \emph{rewards} are the \emph{acknowledgment} (\emph{Ack}) signals received after transmitting packets to the gateway. In this way, a successful transmission is considered when an acknowledgment is received, and a learning approach is employed to select the best channel.

We address the problem of channel selection taking into account the described Markov model for the retransmissions of end-devices. It motivates our present work for which we consider the retransmissions in the analysis of MAB algorithms. 

%

\section{Motivations for the proposed approach} \label{sec:motivations}

When a device experiments a collision, it goes in a back-off state to retransmit the same packet on a channel.
If all devices remain in the same channel for retransmissions, it could result in a sequence of successive collisions with the same devices' packets that previously collided.
Thus, it seems interesting to consider in the decision making policy the possibility for a device to retransmit in a different channel.
One of our motivations to develop new MAB algorithms for our problem is this option of using a different communication channels between the first transmission and the next retransmissions.

By considering this possibility, the device will have to learn more, thus, we expect the learning time to be longer, but it could be possible that the final performance gain (\ie, in terms of successful transmission rate) increases too.
The next Section~\ref{sec:numExp} presents analysis to check this performance gain, for various heuristics based on the \UCB{} algorithm.

Here after, we start by presenting a mathematical derivation that backups this idea.
To do so, we study the collision probabilities considering the Markov process depicted in Fig.~\ref{fig:Markov_model}, and foresee the impact of addressing bandit strategies, as well as setting guidelines for the design of heuristic approaches.

\subsection{Probability of collision at the second transmission slot}

As it is well known, having a collision during an access time can be overcome by a retransmission procedure (this can take several retransmission attempts).
What interest us here, is to obtain a mathematical approximation of the collision probability at the second transmission slot $p_{c1}$, as a function of the first collision probability $p_{c}$.

We consider two hypotheses $\mathcal{H}_{1}$ and $\mathcal{H}_{2}$ defined as,
\begin{itemize}
	\item $\mathcal{H}_{1}$:
    The probability $p_{c1}$, is composed by the sum of two probabilities: i)
    the probability of colliding consecutively twice, \ie, the devices that collide at a given time slot and collide again when retransmitting their packets,
    and ii) the probability of collision among devices that did not collide in the same previous collision. Moreover, we suppose that the number of devices involved in a collision is small in comparison to the total number of devices.
	\item $\mathcal{H}_{2}$:
	The total number of the back-off stages at time $t$ is constant, and it is assumed to be large enough to consider that no device will ever be in the last failure state (this case is the one on the right side in Figure~\ref{fig:Markov_model}), after $M$ successive failed retransmissions.
\end{itemize}

Considering one device and a channel,
we denote $x_t^i$ the probability that it is transmitting a packet for the $i+1$ time in a given time slot $t$ (with $i\in \llbracket 0, M-1 \rrbracket$),
and let $x_t = \sum_{i=0}^{M-1}x_t^i$ be the probability that it transmits a packet.
We consider $N$ active devices following the same policy.

%

We assume to be in the steady state \cite{Norris98}, in our Markov chain model depicted in Figure~\ref{fig:Markov_model}, and thus the probabilities no longer depend on the slot number $t$ (\ie, $\forall t, x_t=x$).
Therefore, the probability that this device has a collision at the first transmission is $p_c$, and has the following expression
\begin{equation}
	p_c = 1-\left(1-x\right)^{N-1} \iff x = 1-\left(1-p_c\right)^{\frac{1}{N-1}}. \label{eqn:1}
\end{equation}

Moreover, from \eqref{eqn:1} we define the probability $p_{cp}(n)$ that involves the collision of $n$ packets sent by each IoT device (for any $1\leq n \leq N-1$), during the first transmission slot, and is defined by the following equation
\begin{equation*}
	p_{cp}(n) = {N-1 \choose n} \; x^n \left(1-x\right)^{N-1-n}. \label{eqn:2}
\end{equation*}

As explained above, if an IoT device experiences a collision at the first transmission, it proceeds for the retransmission of its packet after a random back-off interval.
We denote $p_{ca}$ the probability to have a collision with a packet involved in the previous collision.
Under the $\mathcal{H}_{1}$ assumption, the number of packets involved in the same previous collision remains very small in comparison to the total number of devices that may transmit during this time. In other words, this collision probability does not depend on previous retransmissions and is equal to $p_c$.
So, the probability that the same device's packet experiences again a collision at the second time slot is
\begin{equation}
\label{eq:decomppc1}
p_{c1} = p_{ca}+\left(1-p_{ca} \right)p_c.
\end{equation}

If the device has a collision at the first attempt, we consider $p_{bp}(n)$ the probability that it has a collision with \emph{exactly} $n$ packets (for any $1\leq n \leq N-1$), and that \emph{at least one} of the $n$ devices involved in this first collision chooses the same back-off interval,

\begin{equation}
    p_{bp}(n) = {N-1 \choose n} x^n \left(1-x\right)^{N-1-n}\left[1-\left( 1-\frac{1}{m}\right)^n \right].
\end{equation}

Besides, $p_{ca}$ is the conditional probability of collision with a packet sent by a device involved in the previous collision given that the packet experienced collision at its first transmission.
Hence, under hypothesis $\mathcal{H}_{2}$, we can use Bayes theorem and the law of total probability to relate $p_{ca}$ with $p_{bp}(n)$, and the different probabilities that a device experienced a collision during the first slot and has the same back-off interval for its retransmission is,
%
\begin{equation}\label{eq:sumpca}
	p_{ca} = \frac{1}{p_c}\sum_{n=1}^{N-1} p_{bp}(n).
\end{equation}

\begin{figure}[htp!]  
	\centering
	\includegraphics[width=1.00\linewidth]{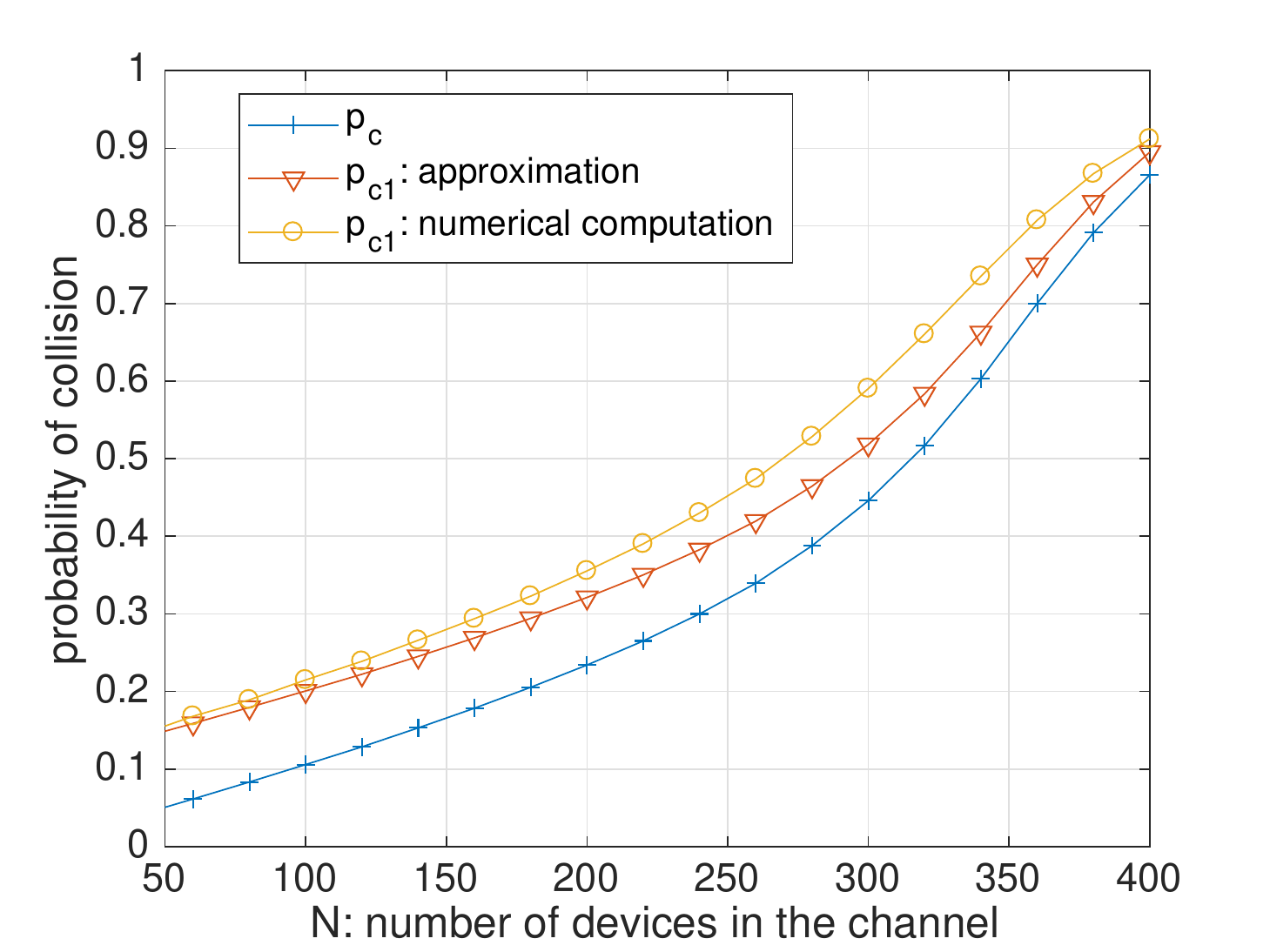}
	\caption{Our proposed approximation for the probability of collision at the second transmission. It is more precise for smaller values of $N$.}
	\label{fig:Approximation_m10}
\end{figure}

Therefore, the expression of $p_{ca}$ is
\begin{align}\label{eq:sumpc2}
	& \frac{1}{p_c} \sum_{n=1}^{N-1}{N-1 \choose n} x^n \left(1-x\right)^{N-1-n}\left[1-\left( 1-\frac{1}{m}\right)^n \right]\nonumber \\
	& = 1- \frac{1}{p_c}\sum_{n=1}^{N-1}{N-1 \choose n} x^n \left(1-x\right)^{N-1-n}\left( 1-\frac{1}{m}\right)^n.
\end{align}

Once again under $\mathcal{H}_{1}$, assuming that the number of devices involved in the first collision is small compared to $N-1$, the first $N_0 \ll N-1$ terms of the sum in \eqref{eq:sumpc2} are predominant. We derive,

\begin{equation}\label{eq:sumpca2}
	p_{ca} \simeq  1- \frac{\left(1-x\right)^{N-1}}{p_c}\sum_{n=1}^{N_0}{N-1 \choose n} x^n \left( 1-\frac{1}{m}\right)^n.
\end{equation}

Moreover, for these terms, $n$ is small compared to $N-1$, and so $N-1-n$ can be approximated to $N-1$. Thus it gives,

\begin{equation}\label{eq:sumpca3}
	p_{ca} \simeq  1- \frac{\left(1-x\right)^{N-1}}{p_c}\sum_{n=1}^{N_0}{N-1 \choose n} x^n \left( 1-\frac{1}{m}\right)^n.
\end{equation}

Assuming $\mathcal{H}_{1}$ amounts to consider that $x\ll 1$. As a consequence, the sum in equation \eqref{eq:sumpca3} can be supplemented by negligible terms,

\begin{equation}\label{eq:sumpca4}
	p_{ca} \simeq  1- \frac{\left(1-x\right)^{N-1}}{p_c}\sum_{n=1}^{N-1}{N-1 \choose n} x^n \left( 1-\frac{1}{m}\right)^n.
\end{equation}

We use the binomial theorem to compute the sum in \eqref{eq:sumpca4}, and we rewrite the expression of $p_{ca}$ as
\begin{multline}\label{eq:pca}
	p_{ca} \simeq \frac{1}{p_c}-\\
	\left(\frac{1}{p_c}-1\right)\left[ 1+\left(1-\left(1-p_c\right)^{\frac{1}{N-1}}\right)\left(1-\frac{1}{m}\right)\right]^{N-1}.
\end{multline}

Finally, our approximation of $p_{c1}$ can be obtained by inserting \eqref{eq:pca} in \eqref{eq:decomppc1}.

\subsection{Behaviour analysis of $p_{c}$ and $p_{c1}$}\label{sub:numericalValidationPC1PC}

In order to assess the proposed approximation, we suppose a unique channel where all the devices follow the same contention Markov process.
We simulate an ALOHA protocol with a maximum number of retransmissions $M=10$, a maximum back-off interval $m=10$, and a transmission probability $p=10^{-3}$.
In Fig.~\ref{fig:Approximation_m10}, we show the collision probabilities for different number of devices $N$ (from $N=50$ up-to $N=400$), for both $p_{c}$ and $p_{c1}$.

From this simulations, we can verify that our approximation is very precise for lower values $p_{c1} \leq 30 \%$ (\ie, red and orange curves are quite close).
Moreover, a significant gap between $p_{c1}$ and $p_c$,
of up-to $10\%$, can be observed,
which suggests us to resort to MAB algorithms for the channel selection for both the first transmission and next retransmissions.

\subsection{Learning is useful for non-congested networks}

It is worth to highlight that, if we write \eqref{eq:decomppc1} as $p_{c1} = p_c + p_{ca} \left(1-p_c\right)$,
then it is obvious that $p_{c1}$ is always larger than $p_c$ (as $p_{ca} \left(1-p_c\right) > 0$).
But for large values of $p_c$, $p_{ca}\left(1-p_c\right) \simeq 0$ so the gap gets small,
and for small values of $p_c$ the gap is significant.
Moreover, we can verify (\eg, numerically or by differentiating)
that the gap decreases when $p_c$ increases (for fixed $N$ and $m$).
This backups mathematically the observation we made from Fig.~\ref{fig:Approximation_m10}:
the smaller $p_c$, the larger is the gap between $p_c$ and $p_{c1}$.

We interpret this fact in two different situations.
On one hand, in a congested network, when devices suffer from a large probability of collision on their first transmission (\ie, $p_c$ is not so small), then $p_{c1}\simeq p_c$ and so devices cannot really hope to reduce their collision probabilities even if the use a different channel for retransmission.
On the other hand, if $p_c$ is small enough, \ie, in a network not yet too congested, then our derivation shows that $p_{c1} \gg p_c$, meaning that the possible gain of retransmitting in a different channel that the one used for the first transmission can be large, in terms of collision probability (\eg, up-to $10\%$ in this experimental setting).
In other words, when learning can be useful (small $p_c$), learning to retransmit in a different channel can have a large impact on the global collision rate,
thus justifying our approach.

\section{A well-known MAB Algorithm: \UCB{}}\label{sec:MABalgo}

Without loss of generality, we have adopted a well-studied stochastic MAB learning algorithm, where the reward distributions are unknown and assumed to be independent and identically distributed (i.i.d). The arms model the channels denoted as $C(t) \in \llbracket 1, K \rrbracket$, and the players, the dynamic devices, learn the distributions to be able to progressively focus on the best arm, \ie, the arm with largest mean representing the mean availability of a given channel $k$.

Before presenting our proposed heuristics, we describe a \UCB{} bandit algorithm \cite{Auer}. It has reported to be efficient, while featuring a low complexity for its implementation. For this reason, it has been employed for IoT applications \cite{Bonnefoi17}, and we employ this approach to develop our proposals.

\subsection{The \UCB{} algorithm}\label{sub:algoUCB}

A first approach is to only use an empirical mean estimator of the rewards in every channel, and select the channel with highest estimated mean at every time step; but this greedy approach is known to fail dramatically \cite{Auer02}.
Indeed, with this policy, the selection of arms depends too much on the first draws: if the first transmission in one channel fails and the first one on other channels succeeds, the device will \emph{never} use the first channel again, even if it is the best one (\ie, the most available, in average).

Rather than relying on the empirical mean reward, \UCB{} algorithms instead use a \emph{confidence interval} on the unknown mean $\mu_k$ of each arm, which can be viewed as adding a ``bonus'' exploration to the empirical mean. They follow the ``\emph{optimism-in-face-of-uncertainty}'' principle: at each step, they play according to the best model, as the statistically best possible arm (\ie, the highest \UCB{}) is selected.

More formally, for one device, let $N_k(t)$ be the number of times the channel $k$ (for $k\in \llbracket 1, K \rrbracket$) was selected up-to time $t-1$, for $t\geq 0$
for any $t\in\mathbb{N}$,
\begin{equation}\label{eq:Nkt}
	N_k(t) = \sum_{\tau=0}^{t-1} \mathbbm{1}(C(\tau) = k),
\end{equation}
where $\mathbbm{1}$ is an indicator function that is equal to $1$, if the IoT device chooses, for its $\tau$-th transmission, the channel $k$, and $0$ otherwise. The empirical mean estimator $\widehat{\mu_k}(t)$ of channel $k$ is defined as the mean reward obtained up-to time $t-1$,
\begin{equation}\label{eq:mukt}
	\widehat{\mu_k}(t) = \frac{1}{N_k(t)} \sum_{\tau=0}^{t-1} r_k(\tau) \mathbbm{1}(C(\tau) = k).
\end{equation}
where $r_{k}(t)$ is the reward obtained after transmission in channel $k$ at time $t$ ($1$ for a successful transmission, and $0$ otherwise)
A \emph{confidence} term $B_k(t)$ is given by \cite{Auer02},
\begin{equation}\label{eq:Bkt}
	B_k(t) = \sqrt{\alpha \log(t) / N_k(t)},
\end{equation}
where $\alpha$ refers to an exploration coefficient\footnote{~In fact, the larger this coefficient is, the longer the exploration, while the \UCB{} algorithm is proven to be order optimal for $\alpha>0.5$ \cite{Bubeck12}, and has reported a good performance for lower values of $\alpha>0$.},
that we chose equal to $1/2$, as suggested in \cite{Audibert07} and as done in previous works \cite{Bonnefoi18,Bonnefoi17}.
Then, an upper confidence bound in each channel $k$ is defined as
\begin{equation}\label{eq:ucb}
	U_k(t) = \widehat{\mu_k}(t) + B_k(t).
\end{equation}
Finally, the transmission channel at time step $t$
is the one maximizing this \UCB{} index $U_k(t)$,
as it is the one expected to be the best one at the current time step $t$,
\begin{equation}\label{eq:maxucb}
	C(t) = \arg\max_{1\leq k \leq K} U_k(t).
\end{equation}

The \UCB{} algorithm is implemented independently by each device, and we present it in Algorithm~\ref{algo:UCB}.
Note that a device using this first approach is only able to select a channel for the first and all the corresponding retransmissions of a packet.

\vspace*{-10pt}
\begin{small}
\begin{figure}[h!]
	\centering
	\removelatexerror
	\begin{algorithm}[H]
		\begin{small}
		\For(){$t = 0, \dots, T$}{
			Compute for each channel $ U_k(t) = \widehat{\mu_k}(t) + B_k(t).$ following Eqs.~\eqref{eq:Nkt},~\eqref{eq:mukt}, and~\eqref{eq:Bkt}\;
			Transmit in channel $C(t) = \arg\max_{1\leq k \leq K} U_k(t)$\;
			Reward $r_{C(t)}(t) = 1$, if \emph{Ack} is received, else $0$\;
		}
		\caption{The \UCB{} algorithm for channel selection.}
		\label{algo:UCB}
		\end{small}
	\end{algorithm}
\end{figure}
\end{small}

\section{Proposed Heuristics}\label{sec:heuristics}

A device that implements the UCB algorithm is led to focus is transmissions and retransmissions in the channel which has been identified as the best. As explained in Section III, focusing in one channel increases the collision probability in retransmissions.
In this Section, we describe the proposed heuristics for the channel selection in a retransmission. It is carried out taking
into account that a device can incorporate a different channel selection strategy while being in a back-off state.
Hence, a natural question is to evaluate whether using this additional contextual information can improve the performance of a learning policy.

For that end, all of our heuristics comprise two stages:
the first stage is a \UCB{} algorithm employed for the first attempt to transmit,
and the second stage is another algorithm used for channel selections for the next retransmissions.

We present below four heuristics for this second stage (short names in ``quotes'' correspond to the legend on Figures~\ref{fig:mainExperiment1}, \ref{fig:mainExperiment2}).

\subsection{Uniform random retransmission (``Random'')}\label{sub:UCBthenRandom}

In this first proposal, the device uses a random channel selection, following a uniform distribution (in $\llbracket 1, K \rrbracket$).
It is described below in Algorithm~\ref{algo:UCBthenRandom}.

\vspace*{-3pt}
\begin{figure}[h!]
	\centering
	\removelatexerror
	\begin{algorithm}[H]
	\begin{small}
	\For(){$t = 0, \dots, T$}{
			\uIf{First packet transmission}{
				Use first-stage \UCB{} as in Algorithm~\ref{algo:UCB}.
			}
			\Else(\tcp*[f]{Random retransmission}){
				Transmit in channel $C(t) \sim \mathcal{U}(1,\ldots,K)$\;
			}
		}
		\caption{Uniform random retransmission.}    
		\label{algo:UCBthenRandom}
	\end{small}
\end{algorithm}
\end{figure}

\subsection{\UCB{} for retransmission (``Only \UCB{}'')}\label{sub:TwoUCB}

Instead of applying a random channel selection,
another heuristic is to use a second \UCB{} algorithm in the second stage.
In other words, we expect that this algorithm is able to learn the best channel to retransmit a packet.
It is described in Algorithm~\ref{algo:TwoUCB}, and it is still a practical approach, since the storage requirements and time complexity remains linear w.r.t. the number of channels $K$ (\ie, of order $\mathcal{O}(K)$).

Note that, we use the superscript $({}^r)$ to denote the variables
$\widehat{\mu^r_k}(t)$, $B^r_k(t)$ and $U^r_k(t)$,
related to the \UCB{} algorithm employed for the retransmission.

\vspace*{-3pt}
\begin{small}
\begin{figure}[h!]
	\centering
	\removelatexerror
	\begin{algorithm}[H]
		\begin{small}
		\For(){$t = 0, \dots, T$}{
			\uIf{First packet transmission}{
				Use first-stage \UCB{} as in Algorithm~\ref{algo:UCB}.
			}
			\Else(\tcp*[f]{Packet retransmission with $\mathrm{UCB}^r$}){
				Compute for each channel $U^r_k(t) = \widehat{\mu^r_k}(t) + B^r_k(t)$ following Eqs.~\eqref{eq:Nkt},~\eqref{eq:mukt}, and~\eqref{eq:Bkt}\;
				Transmit in channel $C^r(t) = \arg\max_{1\leq k \leq K} U^r_k(t)$\;
				Reward $r^r_{C^r(t)}(t) = 1$, if \emph{Ack} is received, else $0$\;
			}
		}
		\caption{\UCB{} for retransmission.}    
		\label{algo:TwoUCB}
		\end{small}
		\end{algorithm}
\end{figure}
\end{small}

\subsection{$K$ different {\UCB}s for retransmission (``$K$ \UCB{}'')}\label{sub:UCBthenKp1}

Another heuristic is to not use the same algorithm no matter where the collision occurred, but to use $K$ different \UCB{} algorithms.
Meaning that after a failed first transmission in channel $j$, the device relies on the $k$-th algorithm to decide its retransmission.
The corresponding algorithm is depicted in Algorithm~\ref{algo:UCBthenKp1}.
Each of these algorithms are denoted using the superscript $({}^{j})$, for $j\in\llbracket 1, K \rrbracket$.

Although, this approach increases the complexity and storage requirements (of order $\mathcal{O}(K^2)$).
For our LPWA networks of interest, such as LoRaWAN, the cost of its implementation is still affordable, since a small number of channels is used.
For instance, for $K=4$ channels,
the memory to storage $K+1=5$ algorithms is of the order of the requirements to storing one.


\vspace*{-3pt}
\begin{small}
	\begin{figure}[h!]
		\centering
		\removelatexerror
		\begin{algorithm}[H]
			\begin{small}
			\For(\tcp*[f]{At every time step})
			{$t = 0, \dots, T$}{
				\uIf{First packet transmission}{
					Use first-stage \UCB{} as in Algorithm~\ref{algo:UCB}.
				}
				\Else(\tcp*[f]{Packet retransmission with $\mathrm{UCB}^j$}){ 
					j $\leftarrow$ last channel selected by first-stage \UCB\;
					Compute for each channel $U_k^j(t) = \widehat{\mu_k}^j(t) + B_k^j(t)$ following Eqs.~\eqref{eq:Nkt},~\eqref{eq:mukt}, and~\eqref{eq:Bkt}\;
					Transmit in channel $C^j(t) = \arg\max_{1\leq k \leq K} U^j_k(t)$\;
					Reward $r^j_{C^j(t)}(t) = 1$ if \emph{Ack} is received, else $0$\;
				}
			}
			\caption{$K$ different {\UCB}s for retransmission.}
			\label{algo:UCBthenKp1}
			\end{small}
		\end{algorithm}
	\end{figure}
\end{small}

\subsection{Delayed \UCB{} for retransmission (``Delayed \UCB{}'')}\label{sub:UCBwithDelay}

This last heuristic is a composite of
the random retransmission (Algorithm~\ref{algo:UCBthenRandom})
and the \UCB{} retransmission (Algorithm~\ref{algo:TwoUCB}) approaches.
Instead of starting the second stage \UCB{} directly from the first retransmission, we introduce a fixed delay $\Delta\in\mathbb{N}$, $\Delta \geq 1$,
and start to rely on the second stage \UCB{} after $\Delta$ transmissions.
The selection for the first steps is handled with the random retransmission.

The idea behind this delay is to allow the first stage \UCB{} to start learning the best channel, before starting the second stage \UCB{} (see details in Algorithm~\ref{algo:UCBwithDelay}).
The number of transmissions to wait before applying the second algorithm is denoted by $\Delta$, it has to be fixed before-hand.

Note that, we use the superscript $({}^d)$ to denote the variables
related to the delayed second-stage \UCB{} algorithm.


\vspace*{-3pt}
\begin{small}
	\begin{figure}[h!]
		\centering
		\removelatexerror
		\begin{algorithm}[H]
			\begin{small}
			\For(\tcp*[f]{At every time step})
			{$t = 0, \dots, T$}{
				\uIf{First packet transmission}{
			 		Use first-stage \UCB{} as in Algorithm~\ref{algo:UCB}.
				}
				\uElseIf(\tcp*[f]{Random selection}){$t \leq \Delta$}{
					Transmit randomly in a channel $C(t) \sim \mathcal{U}(1,\ldots,K)$.
				}
				\Else(\tcp*[f]{Delayed \UCB{}}){
					Compute for each channel $ U^d_k(t) = \widehat{\mu^d_k}(t) + B^d_k(t)$ following Eqs.~\eqref{eq:Nkt},~\eqref{eq:mukt}, and~\eqref{eq:Bkt}\;
					Transmit in channel $C^d(t) = \arg\max_{1\leq k \leq K} U^d_k(t)$\;
					Reward $r^d_{C^d(t)}(t) = 1$ if \emph{Ack} is received, else $0$\;
				}
			}
			\caption{Delayed \UCB{} for retransmission.}
			\label{algo:UCBwithDelay}
			\end{small}
		\end{algorithm}
	\end{figure}
\end{small}

\section{Simulations to compare our heuristics}\label{sec:numExp}

We simulate our network considering $N$ devices following the contention Markov process described in Section \ref{sec:model}, and a LoRa standard with $K=4$ channels.
Each device is set to transmit with a fixed probability $p=10^{-3}$, \ie, a packet about every $20$ minutes for time slots of $1\;\mathrm{s}$.

For the evaluation of the proposed heuristics, a total number of $T=20 \times 10^{4}$ time slots is considered, and the results are averaged over $10^{3}$ independent random simulations.

\begin{figure}[h!]  
	\centering
		\includegraphics[height=7cm,width=9cm]{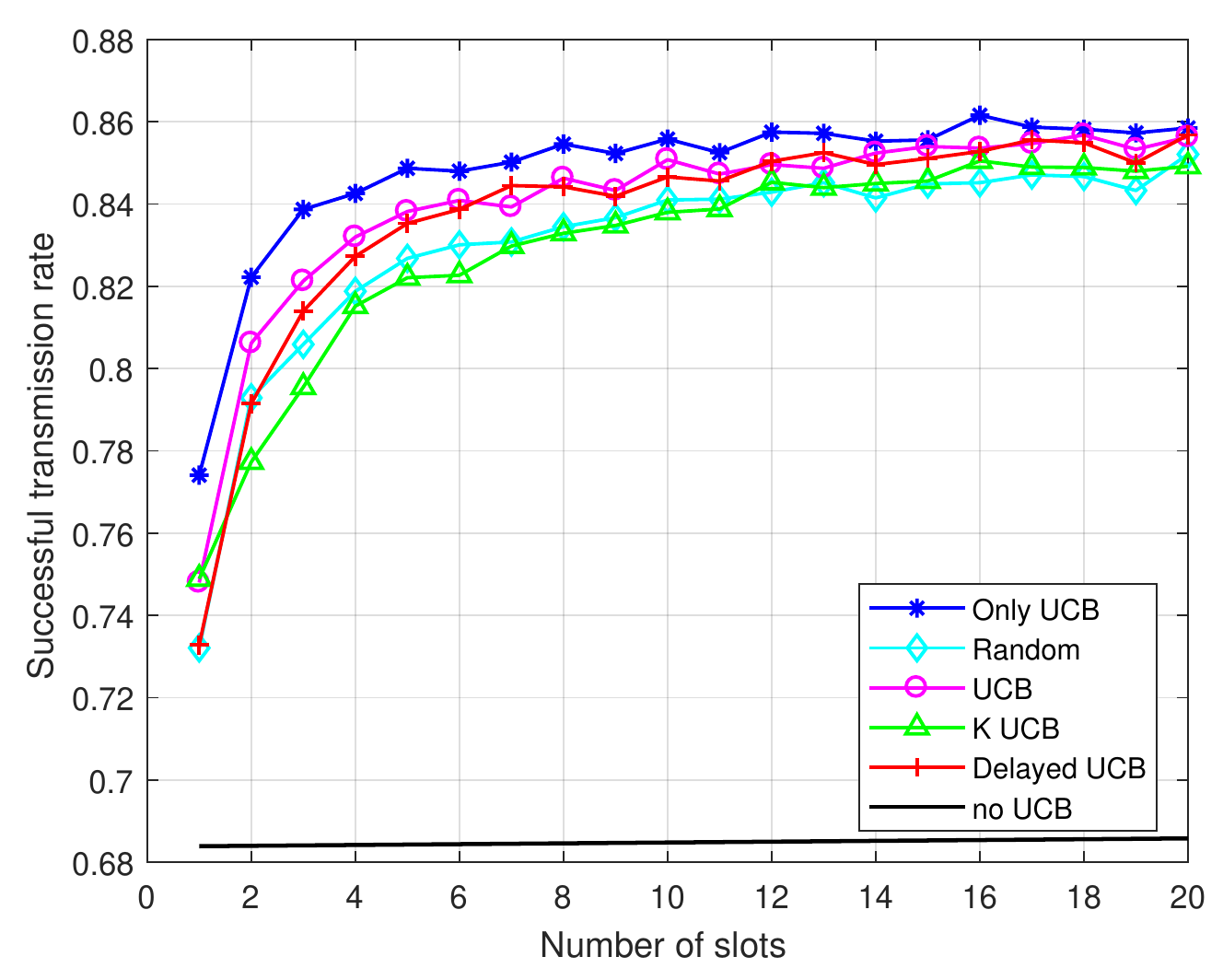}
	  	\caption{
		Comparison among the exposed heuristics for the retransmission: \textcolor{blue}{Only \UCB}, \textcolor{cyan}{Random}, \textcolor{purple}{\UCB}, \textcolor{green}{$K$ \UCB}, and \textcolor{red}{Delayed \UCB}.
		First scenario: learning helps but learning to retransmit smartly is not needed, as we observe that the \textcolor{cyan}{random retransmission} heuristic achieves similar performance than the others.}
	\label{fig:mainExperiment1}
\end{figure}

In a first scenario, we consider a total number of $N=1000$ IoT devices, with a non-uniform repartition of static devices given by $10\%,30\%,30\%,30\%$ for the four channels.
In other words, the channels are occupied $10\%$, $30\%$, $30\%$, and $30\%$ of time, and the contention Markov process considered is given by $M = 5$, and $m=5$.
In Fig.~\ref{fig:mainExperiment1}, we show the successful transmission rate versus the number of slots, for all the proposed heuristics.

A first result is that all the heuristics clearly outperform the non-learning approach that simply use random channel selection for both transmissions and retransmissions (\ie, the \textbf{no \UCB{}} curve).
The improvement of the heuristics over the non-learning approach is evident, and for every heuristic that use a kind of learning mechanism it can be observed a successful transmission rate that increases rapidly (or equivalently an PLR decreasing).
Moreover, all of these approaches show a fast convergence making them suitable for the targeted application.
It is also worths mentioning that the employment of the same \UCB{} algorithm for retransmissions denoted here as ``Only \UCB{}'' achieves the best performance, while a ``Random" retransmission features a slight degradation. This result can be explained as follows: the loss of performance related to the separation of information for several algorithms is greater than the gain obtained by considering the first transmissions and retransmissions separately.

\begin{figure}[h!]  
	\centering
		\includegraphics[height=7cm,width=9cm]{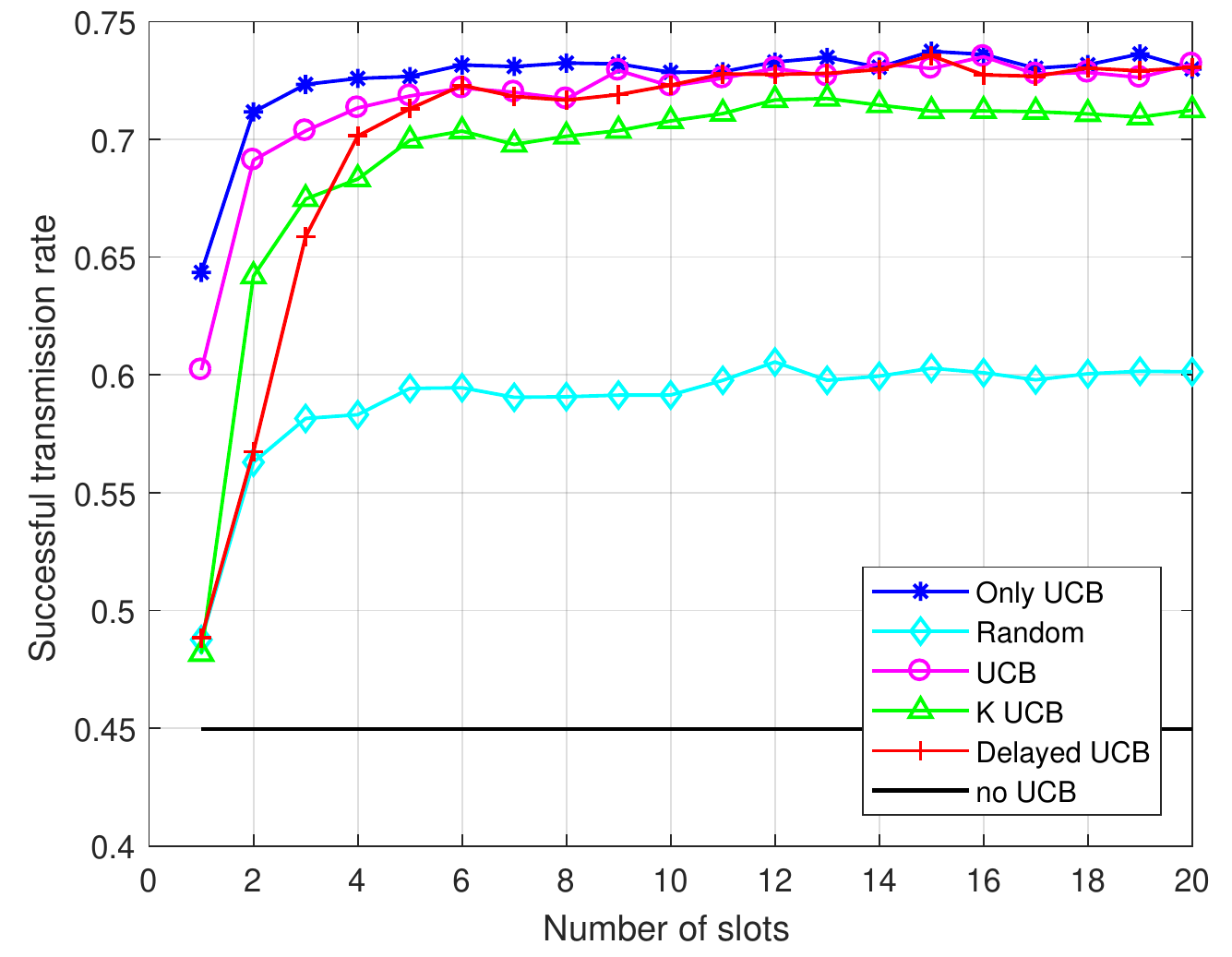}
	  	\caption{Second scenario: learning helps a lot (a gain of $30\%$ in terms of collision probability), and learning to retransmit smartly is needed.
        }
	\label{fig:mainExperiment2}
\end{figure}

We also consider in our analysis the case where $M=5$, and $m=10$ using ALOHA protocol, a statistic distribution of the devices about $40\%, 30\%, 20\%, 10\%$ for the four channels, and $N=2000$ IoT devices.
The corresponding results are depicted in Fig.~\ref{fig:mainExperiment2}.
In this case the successful transmission rate is degraded compared with achieved results in Fig.~\ref{fig:mainExperiment1}, this can be explained with the fact that we are considering in our network more devices that increase the collision probability.
It is important to highlight, that the ``Random" retransmission heuristic shows a poor performance in comparison to the other heuristics, and it can be attributed to the fact that the number of retransmission is increased, and consequently a
learning approach is able to take advantage of it.
Furthermore, the ``\UCB'', ``$K$ \UCB'' and ``Delayed \UCB'' heuristics behave similarly than ``Only \UCB'', after a similar convergence time.
%

The conclusions we can draw from depicted results are twofold.
First, MAB learning algorithms are very useful to reduce the collision rate in LPWA networks, a gain of up-to $30\%$ of successful transmission rate is observed after convergence.
A second conclusion that can be highlighted is that, using learning mechanisms for retransmissions can be an interesting way to reduce collisions in networks with massive deployments of IoT as this can be checked in Fig.~\ref{fig:mainExperiment2}, where the random retransmission heuristic is not very advantageous in front of the  \UCB-based approaches that use learning for channel selection during the retransmission procedure.

\section{Conclusions}\label{sec:conclusion}

In this paper, we presented a retransmission model of LPWA networks based on an ALOHA protocol, slotted both in time and frequency, in which dynamic IoT devices can use machine learning algorithms, to improve their PLR when accessing the network.
The main novelty of this model is to address the packet retransmissions upon radio collision, by using a Multi-Armed Bandit framework.
We presented and evaluated several learning heuristic that try to learn how to transmit and retransmit in a smarter way, by using the \UCB{} algorithm for channel selection for first transmission, and different proposals based on \UCB{} for the retransmissions upon collisions.

We showed that incorporating learning for the transmission is needed to achieve optimal performance, with significant gain in terms of successful transmission rate in networks with a large number of devices (up-to $30\%$ in the example network).
Our empirical simulations show that each of our proposed heuristic outperforms a naive random access scheme.
Surprisingly, the main take-away message is that a simple \UCB{} learning approach, that retransmit in the same channel, turns out to perform as well as more complicated heuristics.

\subsection*{Future works}
The utility and impact of the proposed approaches for LPWA networks motivates us to address several subjects as future works. Among them, the non-stationarity of the channel occupancy caused by the learning policy employed by the IoT devices.
For that end, modifications of MAB algorithms have been proposed, such as Sliding-Window-\UCB{} or Discounted-\UCB{} \cite{Garivier08}
or more recently M-\UCB{} \cite{CaoZhenKvetonXie18},
that nevertheless have not been explored for the targeted problem.

In order to validate our results in a realistic experimental setting and not only with simulations, future works include a hardware implementation of the analyzed models to complete our recent works \cite{modiDemo2016,Besson2019WCNC}.
A hardware demonstrator could be also benefit to study other settings by removing some hypotheses, for instance by studying a similar model in non-slotted time.

\subsection*{Note on the simulation code}
The source code (MATLAB or Octave) used for the simulations and the figures is open-sourced under the MIT License, at \texttt{Bitbucket.org/scee\_ietr/ucb\_smart\_retrans}.

\bibliographystyle{ieeetr}
\bibliography{IEEE_WCNC__2019__Paper__BMBM}

\end{document}